\documentclass[reprint,superscriptaddress,aps,prx]{revtex4-2}
\usepackage{amsmath,epsfig,amssymb,subfigure,bm,dsfont}
\usepackage{lipsum} 

\usepackage{color}
\usepackage{graphicx}
\usepackage{dcolumn}
\usepackage{bm}
\usepackage{bbold}
\usepackage{upgreek}
\usepackage{amsfonts,color}
\usepackage{amsmath}
\usepackage[unicode=true, breaklinks=false, pdfborder={0 0 1}, backref=false, colorlinks=true, linkcolor=blue, urlcolor=blue, citecolor=blue]{hyperref}

\newcounter{lastnote}

\begin{document}
\title{Dynamical Stabilization of Inverted Magnetization and Antimagnons by Spin Injection in an Extended Magnetic System}

\author{Emir Karadža}
\thanks{These authors contributed equally to this work}
\affiliation{%
Department of Materials, ETH Zurich, Zurich 8093, Switzerland
}%

\author{Hanchen Wang}
\thanks{These authors contributed equally to this work}
\affiliation{%
Department of Materials, ETH Zurich, Zurich 8093, Switzerland
}%

\author{Niklas Kercher}
\affiliation{%
Department of Materials, ETH Zurich, Zurich 8093, Switzerland
}%

\author{Paul No\"el}
\affiliation{%
Department of Materials, ETH Zurich, Zurich 8093, Switzerland
}%
\affiliation{Université de Strasbourg, CNRS, Institut de Physique et Chimie des Matériaux de Strasbourg, UMR 7504, Strasbourg F-67000, France}

\author{William~Legrand}
\affiliation{%
Department of Materials, ETH Zurich, Zurich 8093, Switzerland
}%
\affiliation{%
Universit\'e Grenoble Alpes, CNRS, Institut N\'eel, Grenoble 38042, France
}%

\author{Richard Schlitz}
\email{richard.schlitz@uni-konstanz.de}
\affiliation{%
Department of Materials, ETH Zurich, Zurich 8093, Switzerland
}%
\affiliation{Department of Physics, University of Konstanz, 78457 Konstanz, Germany}

\author{Pietro Gambardella}
\email{pietro.gambardella@mat.ethz.ch}
\affiliation{%
Department of Materials, ETH Zurich, Zurich 8093, Switzerland
}%

\date{\today}

\begin{abstract}
Dynamical perturbations can modify the energy landscape of a physical system, such that unstable equilibrium configurations become stable when subject to an external drive. The magnetic analog of such dynamical stabilization corresponds to saturation of the magnetization against an external field. Here we report dynamical stabilization of the magnetization in thin film bismuth-substituted yttrium iron garnet by spin current injection from an adjacent Pt layer. 
Magneto-optical Kerr effect measurements demonstrate magnetization reversal against magnetic fields up to 3000 times larger than the film’s coercivity once the spin injection surpasses a critical threshold associated with negative damping. Micromagnetic simulations reveal that this process is mediated by the excitation of a large population of incoherent magnons with non-zero wave vector, leading to a transient shortening and subsequent stabilization of the inverted magnetization. The elementary excitations of the high-energy inverted magnetization state are shown to be antimagnons, quasi-particles carrying opposite energy and spin relative to magnons. Our results further reveal how the system's size and minimization of nonlinear magnon scattering processes play a key role in dynamical stabilization, opening new avenues for magnetic state control beyond conventional magnetization switching.  
Dissipation-driven phase transitions in large-area magnetic systems provide a solid-state platform to study magnonic analogs of relativistic phenomena, such as Klein tunneling and black holes, as well as spin-wave amplification and lasing.
\end{abstract}

\maketitle

\section{Introduction}

Magnons are the quantized excitations of a magnetically ordered material. Their bosonic character gives rise to collective phenomena inaccessible to fermionic systems, such as nonlinear dynamics~\cite{Rezende1990, Pirro2021}, Bose-Einstein condensation~\cite{Demokritov2006, Kleinherbers2025}, steady-state auto-oscillations~\cite{Collet2016,Demidov2020}, and long-range transport of coherent and incoherent angular momentum~\cite{Demidov2017,Cornelissen2015,Goennenwein2015,Pirro2021}. 
These features not only establish magnons as a leading candidate for energy-efficient information technologies~\cite{Kruglyak2010,Chumak2015,Chumak2022} but also highlight their potential as an ideal platform to study bosonic transport~\cite{Cornelissen2016, Wimmer2019, Schlitz2021, Kohno2023, Kohno2023_2}.

\begin{figure*}[t]
\hspace{-0.5cm}\includegraphics[width=175mm]{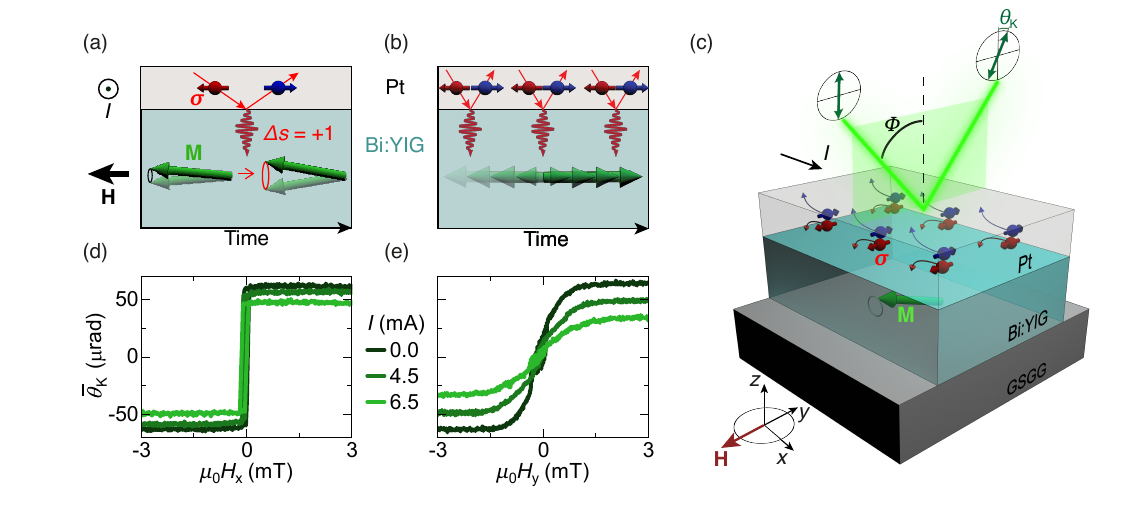}
\caption{Schematic of the experiment and MOKE measurements.
(a) An electric current $I$ flowing in Pt generates an in-plane spin accumulation at the Pt/Bi:YIG interface via the spin Hall effect. Angular momentum is then injected into the Bi:YIG film via spin-flip scattering, resulting in magnon creation when the spin moment $\bm{\upsigma} \parallel \mathbf{M}$ (i.e., spin magnetic moment antiparallel to $\mathbf{M}$).
(b) In-plane reorientation of $\mathbf{M}$ driven by strong spin injection from Pt, which generates a large population of magnons.
(c) Illustration of the longitudinal MOKE geometry, spin injection, and Bi:YIG/Pt bilayer grown on a GSGG substrate.
(d,e) Field-dependent MOKE measurements under different alternating currents with  $\mathbf{H}$ applied along $\mathbf{x}$ and $\mathbf{y}$, respectively. $\bar{\theta}_{\mathrm{K}}$ is the time-average of the Kerr rotation during current injection. 
}
\label{fig1}
\end{figure*}

Recently, attention has shifted to the properties of highly nonequilibrium magnonic systems, with a key theoretical prediction being the existence of antimagnons~\cite{Harms2021, Harms2022, Harms2024, Adorno2024, Kleinherbers2025}, spin-wave modes with left-handed precession that carry opposite spin relative to right-handed conventional magnons. Unlike magnons, whose excitation increases the energy of the magnetic system relative to the ground state, antimagnons are quasi-particle excitations that lower the energy of the system, existing only when the magnetization $\mathbf{M}$ is dynamically stabilized near an energy maximum~\cite{Harms2024}. This situation corresponds to the counterintuitive state where $\mathbf{M}$ is saturated opposite to an external magnetic field $\mathbf{H}$. Despite growing theoretical interest~\cite{Harms2021, Harms2022, Harms2024, Adorno2024, Kleinherbers2025}, experimental evidence for such a dynamically stabilized phase that can host antimagnons is scarce.

Early studies have demonstrated nanoscale inverted magnetization domains driven by spin transfer torque in spin valve devices, characterized by macrospin or soliton dynamics~\cite{Sun2000, zyilmaz2003,Mohseni2013}. However, achieving uniform dynamical stabilization of the magnetization in a macroscopic system is challenging. It requires driving a phase transition in an open system that continuously exchanges energy with an overwhelming number of dissipation channels, primarily arising from magnon-magnon scattering. If dissipation is too strong, the system simply relaxes to the low-energy equilibrium state. If dissipation is too weak, the dynamics remains coherent and close to equilibrium. 
Here we show that injecting a spin current in a macroscopic magnetic layer with minimal damping and nearly compensated magnetic anisotropy can overcome dissipation via magnon scattering. As a result, spin injection generates a large number of magnons across many modes, comparable to the total number of spins, leading to a transient shortening of the net magnetization and its reemergence in a persistent inverted state. The dynamical stabilization of $\mathbf{M}$ against $\mathbf{H}$ leads to the emergence of antimagnons as the excitation quanta of the inverted state. Such a dissipative phase transition, involving finite-wavevector magnons and collective instabilities, is fundamentally distinct from coherent rotational or domain-wall-based magnetization reversal induced by a magnetic field or spin torque, in which the local magnetization retains a constant magnitude~\cite{Ralph2008,Manchon2019, Sun2000, Berkov2008, Baumgartner2017}. Dynamical stabilization thereby introduces a powerful mechanism to manipulate both the magnetization state and the excitation spectrum of extended magnetic systems.

\section{Experiment}

We utilize a $50$~$\upmu\mathrm{m}\times5$~$\upmu\mathrm{m}$, 10-nm-thick bismuth-substituted yttrium iron garnet (Bi:YIG) as magnetic layer and a 5-nm-thick Pt overlayer as spin source (see Appendix~\ref{methods:sample}). Pt features strong spin-orbit coupling and enables charge-to-spin current conversion via the spin Hall effect~\cite{Demidov2011, Sinova2015, Demidov2017}. 
Bi:YIG is a magnetic insulator that offers low magnon dissipation~\cite{Soumah2018}, enhanced magneto-optical properties~\cite{Hansen1983}, and most importantly, tunable magnetic anisotropy~\cite{Wang2025growth}. Elliptical magnetization precession due to shape anisotropy of magnetic films gives rise to nonlinear magnon-magnon scattering. Compensating this shape anisotropy to obtain circular magnetization precession is thus essential to minimize nonlinear magnon scattering, which otherwise limits the maximum number of magnons~\cite{DeLoubens2005,Demidov2020, Pirro2021}.
Driving a charge current in Pt along the $\mathbf{x}$ direction generates a spin accumulation at the interface with Bi:YIG with spin moment $\bm{\sigma}\parallel -\mathbf{y}$, inducing magnon excitation (annihilation) via spin-flip scattering when $\bm{\sigma}$ is parallel (antiparallel) to $\mathbf{M}$~\cite{Takahashi2010} [Fig.~\ref{fig1}(a)]. For a sufficiently large and sustained current leading to overcritical magnon creation, the magnetization of Bi:YIG gradually inverts, resulting in a switched state with $\mathbf{M}\parallel -\mathbf{H}$ [Fig.~\ref{fig1}(b)].
To probe the magnetization response to spin injection, we measure the longitudinal magneto-optical Kerr effect (MOKE) as a function of applied magnetic field and current. In this geometry, the rotation of the reflected light’s polarization, $\theta_\mathrm{K}$, is directly proportional to the in-plane magnetization component $M_{\mathrm{y}}$ [Fig.~\ref{fig1}(c)].
 
We first characterize how the time-averaged Kerr response $\bar{\theta}_{\mathrm{K}}$ is affected by an alternating current in Pt to gauge the impact of magnonic effects. Figure~\ref{fig1}(d) shows $\bar{\theta}_{\mathrm{K}}$ of Bi:YIG/Pt (see Appendix~\ref{methods:MOKE}) when $\mathbf{H}$ is varied along the current direction $\mathbf{x}$.
For zero current, a square hysteresis loop is observed with coercivity $(0.07~\pm~0.02)$~mT. Upon increasing the current amplitude, $\theta_{\mathrm{K}}$ is reduced by Joule heating, which reduces the saturation magnetization of Bi:YIG [see Supplemental Material (SM)~\cite{SI}].
Applying $\mathbf{H}$ along $\mathbf{y}$, a hard-axis loop with saturation at around 2~mT is observed.
The different hysteresis loops stem from the shape anisotropy induced by patterning the Bi:YIG/Pt device, leading to a preferential alignment of $\mathbf{M}$ along the long Hall bar direction. 
Remarkably, applying $\mathbf{H}$ along $\mathbf{y}$ leads to a more drastic reduction of $\bar{\theta}_{\mathrm{K}}$ than applying $\mathbf{H}$ along $\mathbf{x}$, which indicates nonlinear magnon creation and annihilation by spin injection, since Joule heating is identical for both configurations. The geometry with $\bm{\sigma} \parallel +\mathbf{M}$ ($-\mathbf{M}$) corresponds to the application of a longitudinal spin-orbit torque with antidamping (damping) character~\cite{Manchon2019,Demidov2020}.

\section{Dynamical stabilization of nonequilibrium magnetization}

\begin{figure*}[t]
\hspace{-0.5cm}\includegraphics[width=175mm]{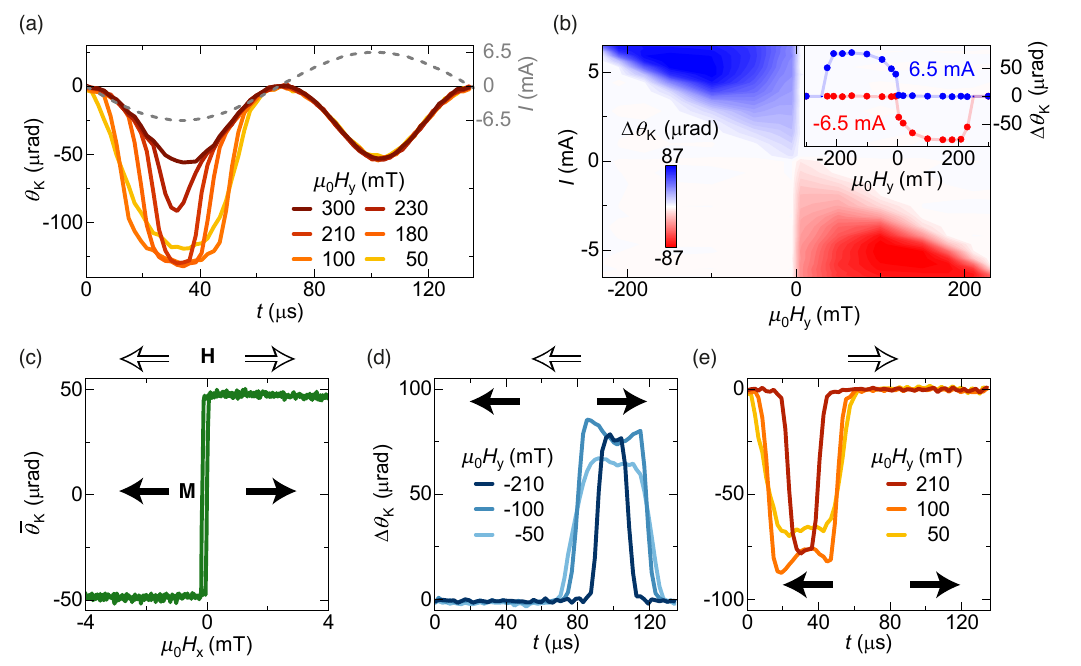}
\caption{Dynamical stabilization of the magnetization by longitudinal spin--orbit torques.
(a) Time-domain Kerr rotation at different external magnetic fields within one period of the alternating current (gray dashed curve).
(b) Maximum excursion of the Kerr rotation per current cycle as a function of applied field and current after subtraction of the Joule heating contribution. Blue and red shading indicate positive and negative current polarity, respectively. The inset shows linecuts at $I=\pm6.5~\mathrm{mA}$, solid lines are guides for the eye.
(c) Reference Kerr rotation measurement showing full reversal of $\mathbf{M}$ in an external field. 
(d,e) Time-domain Kerr rotation traces within one current cycle at $(-50,-100,-210)~\mathrm{mT}$ (d) and $(50, 100, 210)~\mathrm{mT}$ (e). The thermal background due to Joule heating was subtracted from every curve. Arrows indicate the relative orientation between $\mathbf{M}$ and $\mathbf{H}$. Negative (positive) $\mathbf{H}$ favors $\mathbf{M}\parallel -\mathbf{y}$ ($+\mathbf{y}$). Positive (negative) $\Delta\theta_{\mathrm{K}}$ indicates $\mathbf{M}\parallel +\mathbf{y}$ ($-\mathbf{y}$).
}
\label{fig2}
\end{figure*}

To demonstrate the dynamical stabilization of the magnetization, we record $\theta_{\mathrm{K}}$ in the time domain, while applying an alternating current with a peak amplitude of $I= 6.5$~mA (current density 2.6$\times10^{11}$~A/m$^2$ and frequency $\mathrm{f}= 7.32$~kHz. This frequency is very slow compared to the timescale of magnetization dynamics ($\sim$ns), such that $\mathbf{M}$ evolves adiabatically with the sinusoidal current. Figure~\ref{fig2}(a) displays the current trace (dashed line) together with $\theta_{\mathrm{K}}$ at various external magnetic fields $\mathbf{H}\parallel \mathbf{y}$ (solid lines).
The traces show that current injection leads to thermal ($\propto\cos^2(2\pi\mathrm{f} t)$) as well as significant non-thermal effects on $\mathbf{M}$, appearing only in the first half of the current cycle.
These prominent excursions of $\theta_{\mathrm{K}}$ only appear for $\mu_0 H_\mathrm{y}<300$~mT and when the current polarity is such that $\bm{\sigma}\parallel \mathbf{M}$, leading to magnon creation. Figure~\ref{fig2}(b) shows the maximum excursion of the Kerr rotation amplitudes $\Delta \theta_\mathrm{K}$ after subtraction of the cosinus-squared background due to Joule heating. A clear polarity- and field-dependent contrast is observed, confirming that the changes of $\Delta\theta_{\mathrm{K}}$ are driven by magnon creation due to spin injection. 

To relate the excursion of $\theta_\mathrm{K}$ to the magnetization state, we report in Fig.~\ref{fig2}(c) a hysteresis loop that shows full magnetization reversal in an external field, corresponding to a Kerr rotation amplitude of about $95~\upmu\mathrm{rad}$. This measurement is performed with the same current but $\mathbf{H}\parallel \pm \mathbf{x}$ ($\bm{\upsigma} \perp \mathbf{M}$) to prevent magnon creation-annihilation by spin injection. 
Figures~\ref{fig2}(d) and \ref{fig2}(e) show representative time traces of $\Delta\theta_{\mathrm{K}}$ for $\mathbf{H}\parallel -\mathbf{y}$ and $+\mathbf{y}$, respectively.
Direct comparison of these traces with the hysteresis loop in Fig.~\ref{fig2}(c) reveals that the distinct step-like transitions correspond to Kerr rotation amplitudes comparable to those of a full magnetization reversal, especially for the datasets recorded at $\mu_0 H_\mathrm{y} = \pm100~\mathrm{mT}$. 
Inverting $\mathbf{H}$ leads to $\mathbf{M}$ switching for the opposite current polarity, in line with the data in Fig.~\ref{fig2}(b).

These time domain results confirm that, once the current exceeds a threshold value [see Fig.~\ref{fig2}(b)], $\mathbf{M}$ switches and antialigns to a magnetic field as large as $210$~mT. This reversal requires continuous driving via the applied current, indicating a steady-state nonequilibrium phase. Once the current drops below the threshold value, $\mathbf{M}$ realigns with $\mathbf{H}$. Thereby, this switching process is distinct from the conventional remanent switching between two energetically stable states induced by 
either spin transfer torque~\cite{Ralph2008} or spin-orbit torque~\cite{Manchon2019}. Both mechanisms treat the magnetization as a local vector with fixed magnitude~\cite{Sun2000, Berkov2008, Baumgartner2017}, an approximation that breaks down in extended media when a substantial population of nonthermal magnons is excited via longitudinal spin injection~\cite{Chen2024, Nol2025}. Instead, the observed switching corresponds to the dynamical stabilization of $\mathbf{M}$ in an energetically unfavorable state, which becomes effective when the spin current supplies angular momentum faster than the system dissipates it.

\section{Critical threshold and magnon population inversion}

\begin{figure*}[t]
\hspace{-0.5cm}\includegraphics[width=175mm]{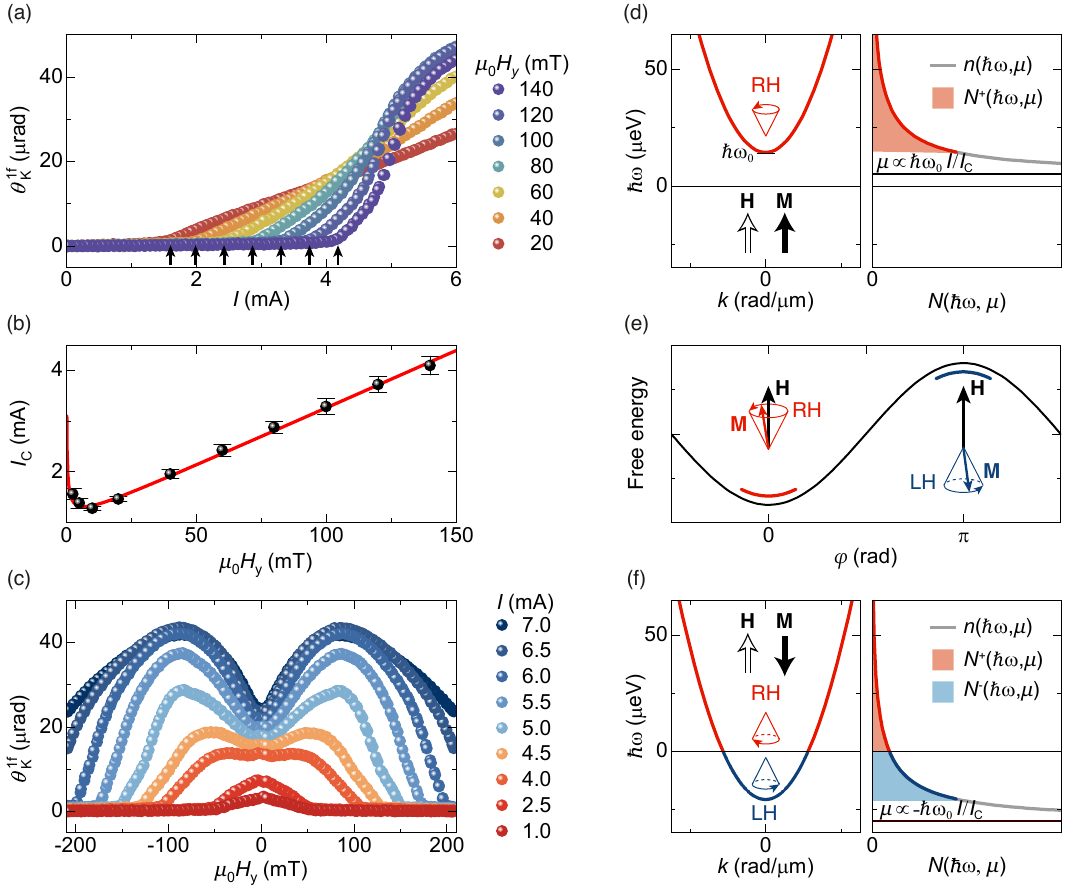}
\caption{Evolution of the dynamic stabilization threshold with magnetic field and current.
(a) Current-dependent harmonic Kerr rotation measurements at different magnetic fields. Black arrows indicate the critical current above which the system transits into the nonlinear regime. 
(b) Critical current vs applied magnetic field. 
(c) Field-dependent harmonic Kerr rotation measurements for different applied currents.
(d) Schematics of thermal magnon dispersion (left) and population (right). The magnon dispersion is parabolic, $\hbar\omega_\mathrm{0}$ indicates the energy of Kittel's mode. The inset shows the magnon chirality relative to $\mathbf{M}$. The right panel shows the Bose-Einstein distribution $n(\hbar \omega, \mu)$ (solid line) for the indicated magnon chemical potential $\mu$ (dashed line); the red shaded area corresponds to the magnon population $N$.
(e) Free energy of the magnetic system. The curvature of the free energy is proportional to the excitation frequency of magnons (red) and antimagnons (blue). $\varphi$ denotes the angle between static $\mathbf{M}$ and $\mathbf{H}$.
(f) Schematics of magnon dispersion (left) and population (right) in the nonequilibrium state. Negative frequency excitations correspond to antimagnons with left-handed chirality. The right panel shows $n(\hbar \omega, \mu)$, $\mu$, and the magnon (red shaded area) and antimagnon (blue-shaded area) populations. 
}
\label{fig3}
\end{figure*}

We now turn to the evolution of the threshold current under different magnetic fields applied along $\mathbf{y}$. To eliminate contributions due to Joule heating and mitigate small drifts of the balanced photodetector, we use an alternating current ($\mathrm{f}=8.75~\mathrm{kHz}$) and demodulate the Kerr rotation at $1\mathrm{f}$ using a lock-in amplifier. The resulting first-harmonic Kerr rotation angle, $\theta_{\mathrm{K}}^{1\mathrm{f}}$, reflects changes of $\mathbf{M}$ that are odd in the applied current.
Figure~\ref{fig3}(a) shows $\theta_{\mathrm{K}}^{1\mathrm{f}}$ as a function of the peak current amplitude. All curves show an abrupt increase beyond a critical current $I_\mathrm{c}$ (indicated by arrows), implying that the switching process is highly nonlinear. Saturation of $\theta_{\mathrm{K}}^{1\mathrm{f}}$ occurs for $\mu_0 H_{\mathrm{y}}>100$~mT at the largest currents, in line with the full reversal of $\mathbf{M}$ reported in Fig.~\ref{fig2}. With decreasing $H_{\mathrm{y}}$, $I_\mathrm{c}$ systematically shifts to lower currents. However, the maximum excursion of $\theta_{\mathrm{K}}^{1\mathrm{f}}$ at small $H_{\mathrm{y}}$ remains markedly lower than at higher $H_{\mathrm{y}}$, even for currents far above $I_\mathrm{c}$.
This observation indicates that full dynamic stabilization of $\mathbf{M}$ can only be achieved at relatively large $\mathbf{H}$ opposing the reversal, a counterintuitive outcome that we will address later.

This incoherent switching process is effectively a dissipative phase transition~\cite{Soriente2021}, which starts when spin injection compensates the positive damping of $\mathbf{M}$ in Bi:YIG.
In a film with vanishing magnetic anisotropy, the intrinsic dissipation rate $\Gamma$ is given by $\Gamma \approx 2 \alpha \gamma \mu_0 H + \Gamma_0$, where $\alpha$ is the Gilbert damping parameter, $\gamma$ is the gyromagnetic ratio and $\Gamma_0$ the decoherence rate due to inhomogeneities in the film~\cite{Collet2016,Wimmer2019,Gckelhorn2021,Kohno2023_2}. Accordingly, $I_\mathrm{c}$ increases linearly with $H_{\mathrm{y}}$ above about 20~mT [Fig.~\ref{fig3}(b)]. Residual shape anisotropy and inhomogeneous broadening result in the upturn of $I_\mathrm{c}$ at lower fields, as reproduced by a fit model including multiple damping contributions (red line, see Appendix~\ref{methods:threshold})~\cite{Gckelhorn2021}.

To obtain further insight into the nonlinear behavior of the system, we performed field-dependent measurements of $\theta_{\mathrm{K}}^{1\mathrm{f}}$ at different currents. The curves in Fig.~\ref{fig3}(c) show that the dynamic stabilization of $\mathbf{M}$ against $\mathbf{H}$ peaks in an intermediate range of fields, consistent with the $I-H$ phase diagram reported in Fig.~\ref{fig2}(b).
Finally, $\theta_{\mathrm{K}}^{1\mathrm{f}}$ vanishes beyond a current-dependent threshold field. This behavior indicates an abrupt suppression of dynamical stabilization in a strong magnetic field, which prevents the reversal of $\mathbf{M}$. 

A schematic model of the magnon energy landscape under incoherent spin injection~\cite{Cornelissen2016, Demidov2020, Kohno2023, Kohno2023_2} captures the main features of dynamic stabilization. We consider a quadratic magnon dispersion $\hbar\omega(k)=\hbar\omega_0 + Ak^2$, where $\omega_0=\gamma\mu_0 H$ is the ferromagnetic resonance frequency, $k$ the magnon wave vector, and $A$ the exchange stiffness, as appropriate for a thin magnetic layer with compensated magnetic anisotropy. At thermal equilibrium, assuming a step-like density of states in two dimensions, the magnon population $N$ is proportional to the Bose-Einstein distribution $n(\hbar\omega, \mu)$, where $\mu$ is the magnon chemical potential [Fig.~\ref{fig3}(d)]. The evolution of $\mu$ with increasing current is described by $\mu = \hbar \omega_0 I/I_\mathrm{c}$~\cite{Kohno2023}, accounting for the progressive buildup of nonequilibrium magnons by spin injection below $I_\mathrm{c}$. Consequently, $N$ gradually increases upon increasing $I$ or reducing $H\propto\omega_0$. When $\mu$ reaches $\hbar\omega_0$, neglecting nonlinear magnon-magnon scattering, $N$ grows exponentially until a new saturated phase with reversed $\mathbf{M}$ is attained~\cite{Harms2024}. This process can be viewed as a population inversion, as all spins within the magnetic layer are flipped into the excited state.

In the dynamically stabilized phase, magnon excitations reduce the total energy by tilting $\mathbf{M}$ away from the unstable equilibrium position [Fig.~\ref{fig3}(e)]. This leads to negative energies of the magnon dispersion for $|k| < \sqrt{H/A}$ [blue arc in Fig.~\ref{fig3}(f)], defining a new class of quantized excitations termed antimagnons~\cite{Harms2021, Harms2022, Harms2024, Adorno2024, Kleinherbers2025}. These antimagnons precess with left-handed chirality relative to $\mathbf{M}$, opposite to conventional magnons. The chemical potential also changes sign: beyond $I_\mathrm{c}$, spin injection reduces the number of antimagnons further stabilizing the reversed phase with $\mathbf{M} \parallel -\mathbf{H}$ by shifting $\mu$ downward, away from the magnon band edge.

While this model explains the nonlinear transition to the inverted phase induced by spin injection, it does not account for the increase of dynamic stabilization against an increasing $\mathbf{H}$, before its suppression in the high field limit [$H_{\mathrm{y}} > 100$~mT, see Figs.~\ref{fig3}(a) and~\ref{fig3}(c)]. This counterintuitive trend reveals an important aspect of dynamic stabilization in real magnetic systems, namely the role played by magnon-magnon scattering. In a realistic setting, the number of nonequilibrium magnons that can fill a particular mode is limited by its nonlinear coupling to other modes. Magnon-magnon relaxation becomes faster with increasing mode occupation~\cite{DeLoubens2005, Kohno2023_2}, such that spin injection becomes less effective in shifting $\mu$ when approaching $I_{\mathrm{c}}$. However, the magnetic field counteracts this effect by enhancing the spatial and temporal coherence of the individual modes, leading to a decrease of the nonlinear coupling between modes with increasing $\mathbf{H}$~\cite{Suhl1959}. Therefore, minimizing unavoidable nonlinear magnon-magnon interactions in a magnetic layer is crucial to induce dynamical stabilization of $\mathbf{M}$. In the present system, this is achieved by compensating the magnetic anisotropy of Bi:YIG and by application of sufficiently strong $\mathbf{H}$ opposing the reversal of $\mathbf{M}$.

\section{Phase transition by incoherent magnon excitation}

\begin{figure*}[t]
\hspace{-0.5cm}\includegraphics[width=175mm]{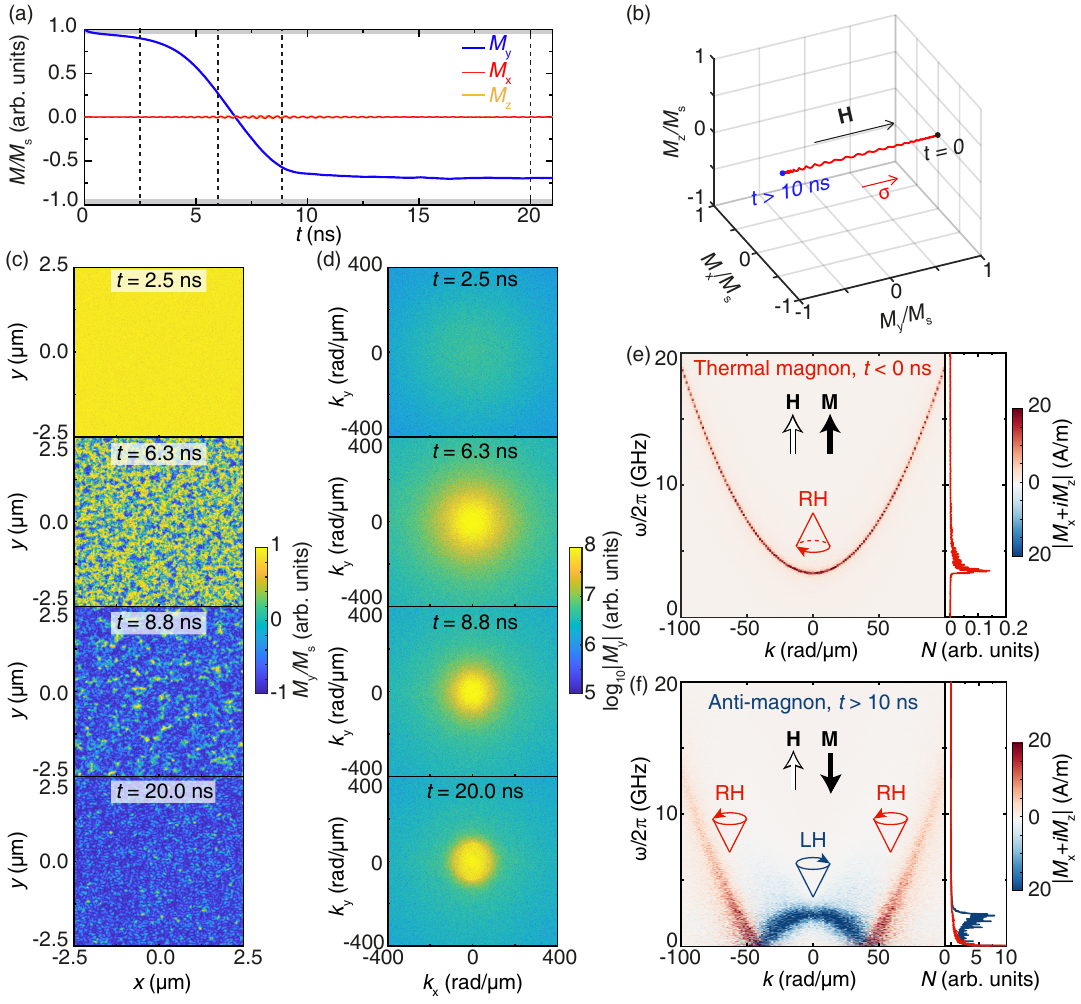}
\caption{
Micromagnetic simulations of the longitudinal magnetization switching and (anti-)magnon distributions.
(a) Time-domain evolution of the spatially averaged magnetization components $M_{\mathrm{x}}$, $M_{\mathrm{y}}$, and $M_{\mathrm{z}}$ during the switching process at $H_{\mathrm{y}}=100$~mT. The curves are shown normalized by the saturation magnetization $M_{\mathrm{s}}$. Gray shaded lines indicate the magnon thermal background at 300~K.
(b) Three-dimensional representation of the switching trajectory.
(c) Spatial maps of $M_{\mathrm{y}}$ at four time snapshots: $t = 2.5$, $6.3$, $8.8$, and $20.0~\mathrm{ns}$.
(d) Wavevector distributions of the magnonic excitations in (c) shown on a logarithmic scale.
(e) Dispersion of thermal magnons and their population (right panel) prior to spin injection. The inset illustrates the chirality of magnons with respect to $\mathbf{M}$.
(f) Dispersion and population of magnons (red) and antimagnons (blue) in the nonequilibrium state. Here the energy of antimagnons is flipped from negative to positive relative to Fig.~\ref{fig3}(f) for comparison with theoretical results~\cite{Harms2024}.
}
\label{fig4}
\end{figure*}

\begin{figure*}[!ht]
\hspace{-0.5cm}\includegraphics[width=175mm]{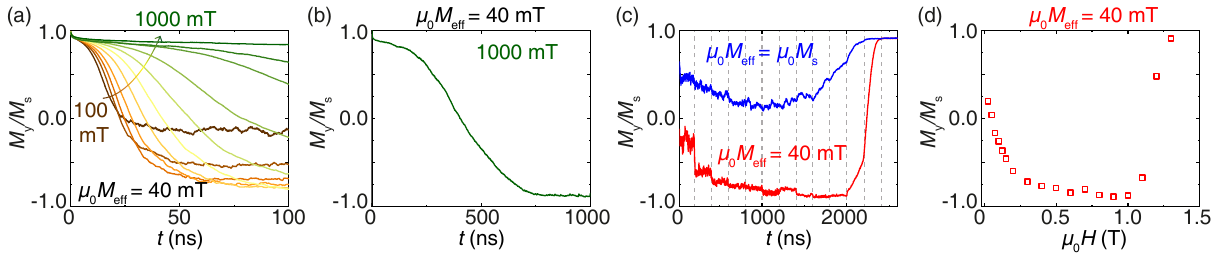}
\caption{
Simulated field dependence of dynamical stabilization.
(a) Time-domain evolution of the spatially averaged $M_{\mathrm{y}}$ under $\mathbf{H}$ applied along $\mathbf{y}$, ranging from 100 to 1000~mT in 100~mT steps. These simulations were performed with a device width of 1~$\upmu$m to reduce computational costs. The applied current density was $0.8\times10^{11}$~A/m$^{2}$. 
(b) Time-domain evolution of the spatially averaged $M_{\mathrm{y}}$ at $H_{\mathrm{y}}=1$~T, showing that the magnetization requires about 800~ns to stabilize against a high field due to enhanced dissipation.
(c) Field dependence of the dynamical stabilization studied by continuously running the simulation and changing the field every 200~ns. Each field configuration evolves from the stabilized state of the previous one, reducing the time required to reach a steady state at high magnetic fields. For $\mu_0H_{\mathrm{y}}>1$~T ($t>2000$~ns), the applied current density is insufficient to stabilize $\mathbf{M}$ opposite to $\mathbf{H}$. The simulations are shown for uncompensated magnetic anisotropy ($\mu_0M_{\rm eff}=\mu_0M_{\mathrm{s}}$, blue) and nearly compensated anisotropy ($\mu_0M_{\rm eff}=40$~mT, red).
(d) Average $M_{\mathrm{y}}$ extracted within the time window where the magnetization is dynamically stabilized by the applied current as a function of magnetic field, showing the same trend as observed experimentally [Fig.~\ref{fig3}(c)].
}
\label{fig5}
\end{figure*}

To obtain a more accurate picture of the switching pathway and magnon-magnon scattering, we employ micromagnetic simulations, modeling spin injection with $\bm{\sigma}\parallel \mathbf{M}$ in a $50$~$\upmu\mathrm{m}\times5$~$\upmu\mathrm{m}$ Bi:YIG layer with compensated magnetic anisotropy in a magnetic field $\mu_0H_{\mathrm{y}}=100$~mT (see Appendix~\ref{methods:simu}). Figure~\ref{fig4} shows the time-domain evolution of the dynamical stabilization process together with the magnon dispersion curves calculated by fast Fourier transforms (FFT) of spatially-resolved magnetization patterns. Upon injecting $\bm{\sigma}$ at $t=0$, the spatially-averaged $M_\mathrm{y}$ component starts to invert, while $M_\mathrm{x}$ and $M_\mathrm{z}$ remain close to zero throughout the process, indicating that switching occurs via shortening of $\mathbf{M}$ rather than through coherent precession. This behavior is further illustrated in Fig.~\ref{fig4}(b), where we report the three-dimensional trajectory of the spatially averaged $\mathbf{M}$ during reversal.

To visualize the phase transition, we present spatial maps of $M_\mathrm{y}$ at four time points, marked by dashed lines in Fig.~\ref{fig4}(a). The first snapshot ($t=$~2.5~ns), shown in Fig.~\ref{fig4}(c), is taken just after thermal magnon stabilization, before significant switching has occurred. The corresponding $k$-space map [Fig.~\ref{fig4}(d)] reveals substantial occupation of nonzero-wavevector states by thermal magnons, with a relatively broad distribution of frequencies [Fig.~\ref{fig4}(e)]. The second snapshot, captured during the inversion process ($t=$~6.3~ns), shows a highly inhomogeneous distribution of $M_\mathrm{y}$ owing to the excitation of numerous nonzero-$k$ magnons~\cite{Ulrichs2020}. The absence of a characteristic length scale is a typical feature of a second-order phase transition near the critical point. The participation of multiple magnons in the transition process is further corroborated by the FFT spectrum, which exhibits a broad region of enhanced intensity. After inversion of $\mathbf{M}$ ($t=$~8.8~ns), magnetic fluctuations are still present, but the FFT indicates a contraction of the excited nonzero-$k$ magnons. Finally, the nonequilibrium steady-state with $\mathbf{M} \parallel -\mathbf{H}$ is stabilized by the continuous injection of $\bm{\sigma}$ after about 10~ns. The last snapshot ($t=$~20~ns) shows that spatial fluctuations of $M_{\mathrm{y}}$ and thus nonzero-$k$ magnons persist in the dynamical stabilization regime. 

To obtain more insight into the magnon properties in this regime, we compare the square of the precessional amplitude ($M_{\mathrm{x}}^2+M_{\mathrm{z}}^2\propto N$, Ref.~\cite{Prabhakar2009}) in the thermal state [Fig.~\ref{fig4}(e)] and switched state [Fig.~\ref{fig4}(f)]. We find that the population of magnons follows the Bose-Einstein distribution and the current-induced shift of $\mu$, in good agreement with the schematic model presented in Figs.~\ref{fig3}(d,f). Note that the negative antimagnon frequencies in Fig.~\ref{fig4}(f) have been flipped relative to Fig.~\ref{fig3}(f) to allow for comparison with theoretical predictions of the magnon-antimagnon dispersion as presented in the literature~\cite{Harms2024}.

Overall, the simulations show that dynamical stabilization occurs via an incoherent process involving the excitation of multiple magnon modes by spin injection until the excess magnon population inverts the magnetization. This process is equivalent to the progressive shortening of $\mathbf{M}$ and its re-emergence along the opposite direction, consistent with our experimental results, which sets it apart from the coherent and domain-wall nucleation-propagation mechanisms characteristic of conventional magnetic field and spin-torque-induced switching~\cite{Ralph2008, Manchon2019,Sun2000, Berkov2008, Baumgartner2017}. 

\begin{figure*}[!t]
\hspace{-0.5cm}\includegraphics[width=175mm]{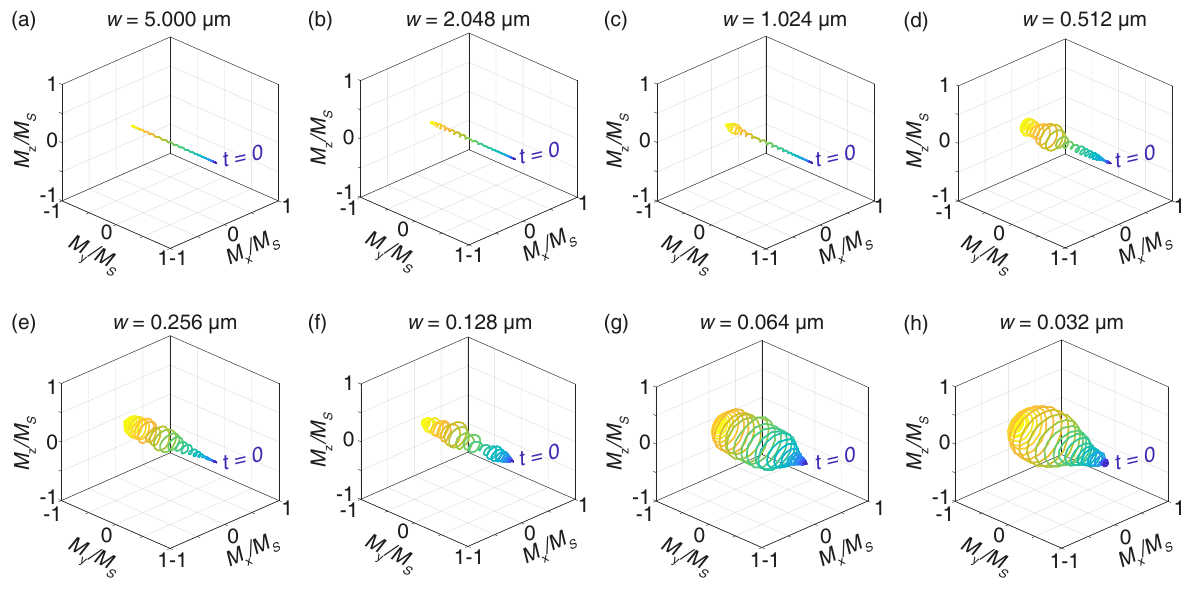}
\caption{
Dynamical stabilization of $\mathbf{M}$ in Bi:YIG layers of decreasing size.
(a-h) Simulated three-dimensional magnetization reversal trajectories in films with widths $w$ of 5.000, 2.048, 1.024, 0.512, 0.256, 0.128, 0.064, and 0.032~$\upmu$m, and lengths equal to ten times the width. A clear transition in reversal behavior is observed, evolving from a linear trajectory to a coherent rotational mode as the film size decreases. This reflects a crossover from a multi-magnon–dominated process in larger systems to a single-magnon–driven switching in smaller systems. The colour of the curve indicates the time from 0~ns (blue) to 25~ns (yellow). A uniform DC current density of $2.6\times10^{11}$ A/m$^{2}$ was applied along $\mathbf{x}$ in all cases.
}
\label{fig6}
\end{figure*}

\section{Field and size dependence}

The strength of the magnetic field is a key factor in stabilizing the inverted state, as shown by the experiments reported in Figs.~\ref{fig3}(a-c). Micromagnetic simulations provide further insight into the stabilization process as a function of field. Figure~\ref{fig5}(a) shows the temporal evolution of the average $M_{\mathrm{y}}$ during spin injection for different field amplitudes opposing switching. With increasing external magnetic field, the time required for the system to reach a stable inverted state becomes longer. For example, for an external field $\mu_0H_{\rm y}=+100$~mT, $M_{\mathrm{y}}/M_{\mathrm{s}}$ converges to approximately $-0.2$ after about $20~\mathrm{ns}$. At $\mu_0H_{\rm y}=+1$~T, the magnetization requires approximately $800~\mathrm{ns}$ to reach the dynamically stabilized state with $M_{\mathrm{y}}/M_{\mathrm{s}}\approx-0.9$, as shown in Fig.~\ref{fig5}(b). This trend reflects the competition between the increased dissipation at higher magnetic fields, which eventually makes the applied current insufficient to drive magnetization reversal, and the suppression of nonlinear magnon–magnon scattering at higher fields, which determines the net value of the stabilized magnetization. The increase of the system's reaction time is in line with a critical slowing down when the dissipation and spin injection rates become equal. 

The dominant nonlinear magnon–magnon scattering arises due to the ellipticity of the magnetization precession, which scales with the magnetic anisotropy parameterized by the effective magnetization $M_\mathrm{eff}$. In our sample, $M_\mathrm{eff}\approx 40~\mathrm{mT}$, such that at low magnetic fields $\mu_0 H_y \lesssim \mu_0 M_\mathrm{eff}$ the magnetization exhibits pronounced elliptical precession due to the thin-film demagnetizing field. As the external field increases, the relative influence of shape anisotropy on the magnon energy is reduced, and the precession becomes more circular~\cite{Bauer2023soft}. In the limit of very large external fields, the magnetization motion approaches purely circular Larmor precession with frequency $\gamma \mu_0 H_{\rm y}/2\pi$, minimizing the energy redistribution between different magnon modes and thereby increasing the efficiency of the dynamical stabilization. 
 
To illustrate these points, we simulated the inversion process for two films with different $M_\mathrm{eff}$. Figure~\ref{fig5}(c) shows the time-dependent evolution of the stabilization level for two cases: one where $\mu_0 M_\mathrm{eff}=40~\mathrm{mT}$, consistent with the film used in the experiment, and another where shape anisotropy dominates and $\mu_0M_{\rm eff}=\mu_0M_{\mathrm{s}}$. Both cases exhibit a characteristic trend in which the stabilization level reaches a peak at specific field values, as summarized in Fig.~\ref{fig5}(d). This behavior is qualitatively consistent with the experimental observations obtained from MOKE measurements in Fig.~\ref{fig3}(c).

Finally, we performed simulations to predict how the reversal dynamics depends on the size of the magnetic system. Figures~\ref{fig6}(a-h) show the trajectory of the magnetization in BiYIG films of constant thickness and lateral sizes ranging from 5~$\upmu$m $\times$ 50~$\upmu$m to 32~nm $\times$ 320~nm, for a fixed current density injected in Pt. As the system size is reduced, the lateral confinement restricts the number of available magnon modes and increases their separation in energy, allowing only modes close to ferromagnetic resonance to be excited. Consequently, the dynamics become increasingly coherent as the size is reduced. In contrast, in larger systems, modes with a broad range of wavelengths can be excited, leading to incoherent dynamics and dynamical stabilization via successive shortening and re-elongation of the magnetization against the external magnetic field. Our simulations thus demonstrate how decreasing the system size drives a crossover from multi-magnon to single-magnon mode dynamics.

\section{Conclusions}\label{sec13}

Steady-state driving by spin injection provides an extremely efficient and robust mechanism for dynamical stabilization of the magnetization, which can be stabilized against external fields up to 3000 times larger than the coercivity of the magnetic layer in the present case. Our results extend the control of spin populations demonstrated in single-spins~\cite{Loth2010} and nanoconstrictions~\cite{zyilmaz2003,Mohseni2013} to macroscopic systems in which the presence of multiple magnon modes and magnon-magnon scattering typically prevent dynamical stabilization. The excitation of a broad spectrum of nonzero-$k$ magnons is a distinctive feature of a dynamical phase transition in extended systems, leading to gradual shortening and reemergence of the magnetization in the anti-aligned state. Reducing dissipation by engineering the magnetic layer's damping and magnetic anisotropy, and application of a strong external field opposing the nonequilibrium magnetization, are thus crucial factors in stabilizing the inverted state.

Future studies could address the efficiency and timescale of dynamical stabilization in systems of varying size, ranging from spin torque nano-oscillators~\cite{Mohseni2013} to magnonic waveguides and devices~\cite{Demidov2017,Kruglyak2010,Chumak2015,Chumak2022,Pirro2021}. Dynamically stabilized magnetization phases have potential applications in spin-wave amplification and lasing~\cite{Doornenbal2019} and open a path toward experimental investigation of deep non-equilibrium magnonics, or \textit{antimagnonics}~\cite{Harms2024}, establishing a solid-state platform to study bosonic analogs of relativistic phenomena, such as the Klein paradox~\cite{Harms2022} and magnonic black holes~\cite{Roldn-Molina2017}. 

From a more general point of view, the dissipative phase transition reported here can be thought of as the magnetic analog of population inversion in a semiconductor laser diode, where photon losses give positive damping and a direct current provides gain. Our work showcases dynamical stabilization by directly controlling dissipation in an open magnetic system instead of the time-dependent modulation of the system's potential energy, as established for mechanical and optical systems~\cite{Bukov2015}.

\begin{acknowledgments}
This research was supported by the Swiss National Science Foundation (Grant No. 200021-236524). H.W. acknowledges the support of the China Scholarship Council (CSC, Grant No. 202206020091). R.S. acknowledges funding by the Deutsche Forschungsgemeinschaft (DFG, German Research Foundation) – project number 425217212. W.L. acknowledges the support of the ETH Zurich Postdoctoral Fellowship Program (21-1 FEL-48). P.N. acknowledges the support of the ETH Zurich Postdoctoral Fellowship Program (19-2 FEL-61). N.K. acknowledges the support of Marie Skłodowska-Curie Grant Agreement N\textsuperscript{\underline{o}} 955671.

\end{acknowledgments}

\section*{Data Availability}

The data that support the findings of this article will be made openly available~\cite{DATASET}.

\clearpage

\appendix

\section{Sample Fabrication}\label{methods:sample}

The Bi:YIG film was grown by radio-frequency magnetron sputtering from a $\mathrm{Bi_{0.8}Y_{2.2}Fe_{5}O_{12}}$ target on a gadolinium scandium gallium garnet (GSGG, $\mathrm{Gd_{3}Sc_{2}Ga_{3}O_{12}}$) substrate. The substrate was heated up to $700~^\circ \mathrm{C}$ for 70 minutes to release adsorbed impurities, in a mixture of Ar and $\mathrm{O_2}$ gases at $0.5~\mathrm{Pa}$. Deposition was performed while maintaining this pressure and temperature at a rate of 0.143 nm/min until a film thickness of $10~\mathrm{nm}$ was obtained, as confirmed by x-ray reflectivity measurements (see SM~\cite{SI}). The film was subsequently annealed for 60 minutes. The sample was kept in the same atmosphere for 120 minutes during cool-down. Finally, a $5 \mathrm{nm}$-thick Pt layer was deposited in-situ on top of the Bi:YIG film by DC magnetron sputtering. The resistivity of the Pt film is $3\times10^{-7}~\mathrm{\Omega m}$ at room temperature. More information on the growth and magnetic properties of Bi:YIG films is presented in Ref.~\cite{Wang2025growth}. The lattice constants of Bi:YIG and GSGG are $12.44 $~\r{A} and $12.55$~\r{A}, respectively. The small lattice mismatch between substrate and Bi:YIG leads to tensile strain, promoting perpendicular magnetic anisotropy~\cite{Soumah2018}. However, capping with the Pt layer induces additional in‑plane anisotropy~\cite{Lee2020, Lee2023, Wang2025}. The saturation magnetization $M_\mathrm{s}$ was estimated to be $100~\mathrm{kA/m}$ (see SM~\cite{SI}). The effective magnetization of the strained Bi:YIG/Pt film is $M_\mathrm{eff}=M_\mathrm{s}-H_\mathrm{PMA}$, where $H_\mathrm{PMA}$ denotes the total perpendicular magnetic anisotropy including the cubic anisotropy contribution.  $\mu_0M_\mathrm{eff}$ of the full film is estimated to be $43~\mathrm{mT}$ (see SM~\cite{SI}).
The Bi:YIG/Pt films were patterned into $5~\mathrm{\upmu m}$ ($10~\mathrm{\upmu m}$) wide and $50~\mathrm{\upmu m}$ ($100~\mathrm{\upmu m}$) long Hall bars, by means of photolithography and Ar$^{+}$ etching. All measurements shown in the main text were performed on a $5~\mathrm{\upmu m}\times 50~\mathrm{\upmu m}$ device at room temperature.

\section{Detection of in-plane magnetization dynamics by longitudinal MOKE}\label{methods:MOKE}

We used a 520~nm laser beam focused through an objective lens (100$\times$ magnification, NA$=0.9$), resulting in a spot size of $\approx 0.5$~$\upmu$m~\cite{Stamm2017}. The polarization of the incident beam was set by a Glan-Thompson polarizer to $s$-polarized light. The rotation of the light polarization due to the change of $\textbf{M}$, the Kerr angle $\theta_\mathrm{K}$, was measured using a balanced photodetector and calibrated using a half-wave plate. The laser beam was focused to the center of the Hall bar. By changing the lateral offset of the laser beam on the entrance pupil of the objective, the incidence angle $\Phi$ can be set to parallel ($0^\circ$) or oblique with respect to the sample's surface normal. To probe the in-plane (longitudinal) evolution of $\textbf{M}$, $\Phi$ was set at $\approx 34^\circ$. The incidence plane was always set parallel to the applied external field $\textbf{H}$, corresponding to the longitudinal MOKE geometry [see Fig.~\ref{fig1}(c)]. Measurements at the inverted incidence angle $\Phi \approx -34^\circ$, allow us to remove any contribution arising from the out-of-plane $\textbf{M}$ component. Therefore, we calculate the Kerr rotation $\theta_{K}$, related to the in-plane $\mathbf{M}$ component as $\theta_\mathrm{K}=(\theta_\mathrm{K(L^+)}-\theta_\mathrm{K(L^-)})/2$, where $+$ and $-$ indicate positive and negative incidence angles, respectively. 

Generally, we resolve both static and dynamic responses of the system to a current-induced stimulus, by applying alternating (sine wave) currents and demodulating the signal from the balanced photodetector using a lock-in amplifier (LIA). Furthermore, we can resolve the Kerr rotation in the time domain, within the cycle of the applied current, by utilizing the built-in scope function of the LIA. The Kerr rotation traces shown in Fig.~\ref{fig2}(a) are averaged over 256 current periods to increase the signal-to-noise ratio. Due to drifts of the balanced photodetector, a current-independent offset was removed from each trace. For the hysteresis loop measurements, this correction was performed by subtracting the mean of each dataset, whereas for the time-resolved measurements, the Kerr rotation corresponding to zero applied current was subtracted from each dataset. The harmonic measurements are unaffected by this issue, as DC offsets of the balanced photodetector are removed by the demodulation.

\section{Threshold for nonlinear magnon excitation}\label{methods:threshold}

Spin injection by the spin Hall effect (SHE) generates a transverse spin current in the Pt layer. Its absorption at the Bi:YIG/Pt interface gives rise to an interfacial spin torque rate rate $\Gamma_\mathrm{ST}$ that counteracts the total (positive) magnetic damping torque rate $\Gamma_\mathrm{D}$ of the Bi:YIG film. Damping compensation occurs when the net torque vanishes, $\Gamma_\mathrm{ST} = \Gamma_\mathrm{D}$,
corresponding to the onset of auto-oscillations~\cite{Collet2016, Wimmer2019, Gckelhorn2021, Kohno2023}.
The spin torque originates from a charge current $I$ flowing through Pt, which generates a spin current density:
\begin{equation}
j_s = \frac{\hbar}{2e}\, \theta_\mathrm{SH}\, \frac{I}{w_\mathrm{Pt} t_\mathrm{Pt}}\, \tanh{\eta},
\qquad \eta = \frac{t_\mathrm{Pt}}{2 l_s},
\end{equation}
where $\hbar$ is the reduced Planck constant, $e$ the elementary charge, and $l_\mathrm{s}$, $t_\mathrm{Pt}$, $w_\mathrm{Pt}$, and $\theta_\mathrm{SH}$ denote the spin diffusion length, thickness, width, and spin Hall angle of Pt, respectively, with $\eta=t_\mathrm{Pt}/(2l_\mathrm{s})$.

The product $l_s \theta_\mathrm{SH}$ therefore represents the efficiency of the SHE spin source: $l_s$ quantifies spin-current transport through Pt, while $\theta_\mathrm{SH}$ converts charge to spin angular momentum. Together they define the ``supply side" of the spin-torque process. The effective damping is therefore~\cite{Collet2016, Wimmer2019, Gckelhorn2021}:
\begin{equation}
\alpha_\mathrm{eff}(H) =
\alpha_\mathrm{G} + \frac{\Delta H_0}{2\sqrt{H(H+M_\mathrm{eff})}},
\end{equation}
where the first term represents the intrinsic Gilbert damping, while the last term accounts for field independent inhomogeneous linewidth broadening $\Delta H_0$  (reflecting the spatial uniformity of the local resonance fields). As the relationship between frequency and field is not linear~\cite{Collet2016}, the damping due to inhomogeneous broadening exhibits a $1/\sqrt{H(H+M_\mathrm{eff})}$ field dependence~\cite{Collet2016, Wimmer2019, Gckelhorn2021}. On the other hand, the dynamic magnetization of Bi:YIG pumps a reciprocal spin current into Pt. This backflow, described by the spin-mixing conductance $g^{\uparrow \downarrow}$, adds an additional damping channel --the spin-pumping contribution $\alpha_\mathrm{sp}$-- to the total magnetic loss rate. The damping due to spin-pumping is proportional to the interfacial spin transparency~\cite{ Zhang2015, Wimmer2019}
\begin{equation}
\mathcal{T}=\frac{g_\mathrm{eff}\tanh{\eta}}{g_\mathrm{eff}+h\sigma_\mathrm{Pt}/(2e^2 l_\mathrm{s})},
\end{equation}
which is quantified by the real part of the spin-mixing conductance~\cite{Tserkovnyak2002a, Zhang2015}:
\begin{equation}
    g_\mathrm{eff} = g^{\uparrow\downarrow}
\frac{h\sigma_\mathrm{Pt}}{2e^2l_s}
\left(g^{\uparrow\downarrow}+\frac{h\sigma_\mathrm{Pt}}{2e^2l_s}\right)^{-1},
\end{equation}
giving
\begin{equation}
\alpha_\mathrm{sp} = 
\frac{\gamma \hbar}{4\pi M_{\mathrm{s}} t_\mathrm{Bi:YIG}} g_\mathrm{eff}.
\end{equation}
Here, $\gamma$ is the gyromagnetic ratio, $\sigma_\mathrm{Pt}$ the conductivity of Pt, and $M_\mathrm{s}$, $M_\mathrm{eff}$, and $t_\mathrm{Bi:YIG}$ denote the saturation magnetization, effective magnetization, and thickness of the Bi:YIG film, respectively. The parameter $g_\mathrm{eff}$ captures the reduction of the bare spin-mixing conductance by spin backflow in Pt, consistent with the reciprocal nature of spin pumping and spin torque~\cite{Collet2016,Kohno2023}.

Balancing the spin-torque rate at the interface $\gamma \hbar j_s /(2 e M_\mathrm{s} t_\mathrm{Bi:YIG})$ with the total damping rate $(\alpha_\mathrm{sp}+\alpha_\mathrm{eff}) \gamma \mu_0 \left( H + M_\mathrm{eff}/2 \right)$, yields the threshold current at which $\mu=\hbar\omega_0$~\cite{Bender2014,Wimmer2019,Gckelhorn2021}:
\begin{align}
I_\mathrm{C}
&= \frac{\sqrt{2}\hbar}{2e}\,
   \frac{\sigma_\mathrm{Pt}}{2 l_s}\,
   \frac{t_\mathrm{Pt} w_\mathrm{Pt}}
        {\theta_\mathrm{SH}\tanh\eta} \notag \\
&\quad \times
   \left(
   1 + 4\pi M_{\mathrm{s}} t_\mathrm{Bi:YIG}\,
   \frac{\alpha_\mathrm{eff}(H)}
        {\hbar \gamma g_\mathrm{eff}}
   \right)
   \gamma \mu_0
   \left( H + \frac{M_\mathrm{eff}}{2} \right).
\label{eq:dampingcomp1}
\end{align}

Here, the factor $\sqrt2/2$ is introduced to account for the conversion from the peak value of the alternating current to its root-mean-square equivalent. 
The ratio $\alpha_\mathrm{eff}/(\hbar \gamma g_\mathrm{eff})$
expresses the competition between the damping to be overcome ($\alpha_\mathrm{eff}$) and the interfacial spin-transfer efficiency ($g_\mathrm{eff}$), analogous to the ratio $\alpha_\mathrm{eff}/\alpha_\mathrm{sp}$ introduced in Refs.~\cite{Collet2016,Kohno2023}: if $\alpha_\mathrm{sp}$ is small (weak coupling), $I_\mathrm{c}$ increases accordingly.

Equation~(\ref{eq:dampingcomp1}) contains both field-independent quantities $l_s$, $\theta_\mathrm{SH}$, and $g^{\uparrow\downarrow}$, which describe the static spin-transport efficiency of the Pt/Bi:YIG interface, and a field-dependent dissipation term $\alpha_\mathrm{eff}(H)$. The explicit field dependence of Eq.~(\ref{eq:dampingcomp1}) arises from three distinct contributions: (i) the FMR linewidth $\Delta\omega(H) \propto \gamma \mu_0 (H+M_\mathrm{eff}/2)$, quantifying the damping rate of magnetization precession; (ii) the inhomogeneous broadening, $\Delta H_0$, arising from sample imperfections whose contribution to the effective damping decreases with increasing field according to $\Delta H_0 / [2 \sqrt{H(H+M_\mathrm{eff})}]$, as higher precession frequencies suppress the relative effect of inhomogeneities; and (iii) the implicit frequency dependence of the intrinsic and spin-pumping damping, $\alpha_\mathrm{G}$ and $\alpha_\mathrm{sp}$. Consequently, the variation of $I_\mathrm{c}$ with magnetic field directly reflects the evolution of the total magnetic damping, whereas the prefactor $(l_\mathrm{s}\theta_\mathrm{SH}g^{\uparrow\downarrow})$ defines the constant efficiency of the spin–current source and interfacial torque transfer.

The critical current reflects a balance between spin-torque injection (proportional to $l_s\theta_\mathrm{SH}$) and damping (scaling with $\alpha_\mathrm{eff}/g_\mathrm{eff}$). Because spin pumping acts as the reciprocal process of the spin-transfer torque, the interface transparency $g^{\uparrow\downarrow}$ governs both how efficiently the magnet pumps spins and how effectively it receives them. The field dependence of $I_\mathrm{c}$ thus directly maps the evolution of $\alpha_\mathrm{eff}(H)$, while the prefactor encodes the static spin-current generation efficiency.

The experimental data in Fig.~\ref{fig3}(b) was fitted using Eq.~(\ref{eq:dampingcomp1}), with $\theta_\mathrm{SH}$ and $g^{\uparrow\downarrow}$ as free parameters while all geometric and magnetic constants were fixed: $\mu_0M_\mathrm{eff}=43~\mathrm{mT}$, $M_\mathrm{s}=100~\mathrm{kA/m}$, $\alpha_\mathrm{G}=0.00125$, $\mu_0\Delta H=0.3~\mathrm{mT}$, $t_\mathrm{Bi:YIG}=10~\mathrm{nm}$, $t_\mathrm{Pt}=5~\mathrm{nm}$, $w_\mathrm{Pt}=5~\mathrm{\upmu m}$, $\sigma_\mathrm{Pt}=3.3\times10^6~\mathrm{S/m}$, $l_\mathrm{s}=2~\mathrm{nm}$ (see SM~\cite{SI}). From the fit we obtain $\theta_\mathrm{SH}=0.073\pm0.002$, $g^{\uparrow\downarrow}=(6.3\pm0.6)\times10^{18}~\mathrm{m^{-2}}$, and estimate $\mathcal{T} = (0.16\pm0.01)$, yielding the effective spin Hall angle $\mathcal{T}\theta_{\mathrm{SH}}=0.012$ used in the micromagnetic simulations.

\section{Micromagnetic Simulations}\label{methods:simu}
All spin relaxation and dynamic processes were simulated using the MuMax$^3$ micromagnetic package~\cite{Vansteenkiste2014}. The simulations were performed for different Hall-bar dimensions; the example shown in Fig.~\ref{fig4} corresponds to a region of $50~\upmu$m $\times$ $5~\upmu$m $\times$ 12~nm ($xyz$) with a unit cell size of $4 \times 4 \times 4$~nm$^3$.

The temperature was fixed at 300~K to account for thermal magnons. The magnetization was initially aligned along the $\mathbf{y}$ axis. We assumed an effective spin Hall angle $\mathcal{T}\theta_{\mathrm{SH}}=0.012$, consistent with measurements of spin-orbit torques on the same devices, and a uniform DC current of 6.5~mA along $\mathbf{x}$, corresponding to a current density of $2.6\times10^{11}$~A/m$^{2}$. The saturation magnetization was set to 100~kA/m. The perpendicular anisotropy was set to 3.75~kJ/m$^{3}$ in order to reproduce an effective magnetization of $\mu_0 M_{\mathrm{eff}} \approx 43$~mT, consistent with the experimental value.

\bibliography{bibliography}

\end{document}